\documentclass[aps,prx,twocolumn,groupedaddress]{revtex4-1}

\usepackage{xcolor}

\usepackage{graphicx}
\usepackage{dcolumn}

\begin{document}



\title{Hydrodynamic instabilities, waves and turbulence in spreading epithelia}


\author{C. Blanch-Mercader and J. Casademunt}
\affiliation{Unitat de Biof\'isica i Bioenginyeria, Facultat de Medicina, Universitat de Barcelona, Barcelona 08036, Spain.}
\affiliation{Laboratoire Physico Chimie Curie, Institut Curie, PSL Research University, CNRS, 26 rue d' Ulm, 75005 Paris, France}
\affiliation{Universitat de Barcelona Institute of Complex Systems (UBICS), Universitat de Barcelona, Barcelona, Spain.}




\date{\today}

\begin{abstract}

We present a hydrodynamic model of spreading epithelial monolayers as polar viscous fluids, with active contractility and traction on the substrate. The combination of both active forces generate an instability that leads to nonlinear traveling waves, which propagate in the direction of polarity with characteristic time scales that depend on contact forces. We show that a viscous fluid model  explains a variety of observations on the slow dynamics of epithelial monolayers, in particular those that had been interpreted before as signatures of elasticity. The non-elastic nature of the waves can be tested on the basis of simple predictions of the model. Our theoretical framework provides new insights such as the interpretation of plithotaxis as a result of a strong flow-polarity coupling, and the quantification of collective force-transmission of cells in terms of the non-locality of interactions. In addition, we study the nonlinear regime of those waves deriving an exact map of the model into the Complex Ginzburg-Landau equation, which provides a complete classification of possible nonlinear scenarios. In particular, we predict the transition to different forms of weak turbulence, which in turn could explain the very unstable and irregular dynamics often observed in epithelia. 
\end{abstract}


\maketitle


\section{Introduction}

In recent years, in vitro epithelial cell monolayers have become a key model system to investigate mechanical aspects of collective cell migration, a generic situation that is directly relevant to a variety of biological processes in living organisms, including morphogenesis  \cite{Lecuit2007,Bryant2008,Friedl2009} or regeneration \cite{Sonnemann2011,Brugues2014,Cochet-Escartin2014a}. In particular, much attention has been focused on the collective mechanisms by which cohesive advancing cell sheets, are capable to transmit and build up intracellular stresses over distances of hundreds of microns, \cite{Leckband2011}. Cells are able to exert actively-driven forces to the substrate underneath and migrate towards the maximum principal stress direction, and simultaneously the instantaneous monolayer stress maps may trigger signaling pathways that affect the mechanical state of individual cells \cite{Roure2005,Trepat2009a}. This suggests a strong interplay between the physical properties of a tissue and the internal structure of the constituting cells, \cite{Weber2011a,Tlili2013}. Consequently, it becomes crucial to develop a solid theoretical framework to interpret the force and kinematic maps nowadays available for spreading epithelial monolayers. Form a physical standpoint, a key point is to elucidate to what extent the phenomena observed, even if strongly regulated biologically, can be tackled in purely mechanical terms. Within this spirit, in recent years, there has been an increasing interest in designing experiments in vitro to probe the mechanics of epithelial tissues in controlled situations \cite{Trepat2009a,Serra-Picamal2012a,Vedula2012,harris2012,Deforet2014,notbohm2016cellular}.

One issue that remains a matter of debate on the theoretical side is whether a continuum description of a tissue must assume a viscous \cite{arciero2011,Lee2011a,Marel2014,recho2016one} or an elastic \cite{Banerjee2012,Kopf2013,wavesMarchetti2015,tlili2016waves} constitutive equation at a given range of time scales of observation, in particular for the long-time regime. This question is nontrivial for living tissues in particular because a given type of cells may respond differently in different environments at the same time scale. For instance, MDCK cells in suspended monolayers under external pulling seem to respond elastically~\cite{harris2012} in time scales for which a freely spreading monolayer on a substrate seems to be flowing like a viscous fluid~\cite{mercader2017effective}. In addition to this intrinsic variations of the mechanical properties of cells in response to the environment, additional confusion may arise when comparing the mechanical response of the tissue to an external force~\cite{harris2012}, to the relationship between stress and strain variables when the stress is autonomously induced by the tissue, implying that the two observables may be related by some additional constraint, either biological or otherwise, that prevents from establishing a direct causal, stimulus-response relationship between both. A clear example of this point is the observation that both stress and strain at the central region of a spreading monolayer have been shown grow linearly with time in some initial range, as a consequence of the fact that the free end of the tissue advances at constant speed. This obviously allows to establish a linear relationship between stress and strain for some time which would be characteristic of an elastic medium, as pointed out in Ref.~\cite{romaric2015}. However, the same time dependent data are consistent with an assumption of a purely viscous constitutive equation, once taken into account that the effective viscosity is time-dependent due to the  narrowing of the cell monolayer as it spreads. This point has been discussed in detail elsewhere \cite{mercader2017effective}.  

In this context, the present study is directly motivated by a series of experiments \cite{Serra-Picamal2012a} on spreading epithelia where ultra-slow elastic-like waves have been reported, in time scales of several hours, where one could argue that, on the basis of the times scales of intracellular processes (around $1$ minute) and cell-cell adhesion kinetics (around $10$ minutes) one should expect viscous behavior. Independent observations of individual displacement of cells and their relative sliding also suggests that the relaxation of stresses is fluid-like. However the same experiments reported a phase lag between stress and strain-rate measurements that is characteristic of elastic waves. The different attempts to model these phenomena so far \cite{Serra-Picamal2012a,wavesMarchetti2015,tlili2016waves}, have assumed an elastic constitutive equation for the tissue and different additional hypothesis to account for the emergence of waves.

In this paper we present a continuum model of a epithelial monolayer spreading on a substrate that is based on a viscous constitutive equation for the medium and combines two sources of active stresses: bulk contractility and traction forces at the contacts with the substrate. We will elucidate a hydrodynamical instability that can explain the emergence of elastic-like waves, in particular in the range of time and length scales of the observation. The mechanism and the physical scenario that accounts for the waves is completely different from that of elastic models, and the waves exhibit distinctive features with no counterpart in those models. In particular we will discuss the 
observations of the experiments from Refs. \cite{Serra-Picamal2012a,Vedula2012,Deforet2014} in the light of our approach. Our model is based on the phenomenological continuum approach of active gels theory \cite{Kruse2005,Julicher2007}, where the medium is treated as polar and the equations are imposed solely by symmetry considerations and linear thermodynamics. Most of the phenomenological parameters of the theory can be estimated from independent experimental observations and those with no direct evidence will be indirectly inferred with the use of the model. In particular the large values obtained for the flow alignment coefficient will provide interesting insights into the phenomenon of plithotaxis. 

In addition to the basic linear instability mechanism for polarized tissues under traction and contraction, we perform a thorough nonlinear study of the problem. We carry out direct numerical simulations deeply in the nonlinear regime, and pursue a weakly nonlinear analysis close to the instability threshold. We derive explicitly the mapping to a 
 Complex Ginzburg-Landau equation, which allows us to classify the different nonlinear dynamical regimes, and in particular to show that the transition to different forms of weak turbulence are generic in our problem. We speculate on the possible relevance of such scenarios as opposed to a purely stochastic origin of the strong fluctuations.

The layout of the paper is as follows. In Section II we present and discuss the continuum model. In Section III, we discuss the linear stability analysis of a homogeneously polarized state, with emphasis on the physical interpretation of the model predictions. In particular, we elucidate a new scenario to explain the emergence of waves in a viscous medium with properties that seem to evoke elasticity. Section IV is devoted to the nonlinear regime, and includes both numerical simulations and the analytical results of the weakly nonlinear theory, with the mapping of the problem into a Complex Ginzburg-Landau equation and the discussion of the transition to different forms of spatio-temporal chaos or weak turbulence. In Section V we revisit some experiments on MDCK cell monolayers, and interpret their results in the light of our theoretical framework. We finally sumarize our results in the concluding section VI. 

\section{Continuum model}

Cells are assumed to have a macroscopic polar order described by the vector field $\mathbf{p}$. At the free-edges of expanding cell sheets, they tend to develop lamelipodium-like structures that require a globally oriented actin cytoskeleton, although cells that are hundreds of microns away from the interface extend basal cryptic lamellipodia underneath the neighboring cells \cite{Farooqui2005a}. In addition, epithelial cells exhibit other type of complex nematic structures, such as stress fibers connecting them. To mimic the tendency of the polarity field to align with the neighbors and thus avoid large gradients, we introduce an effective free energy for these degrees of freedom of the form of the standard free energy of a polar \cite{de1993physics}, 
\begin{equation}
{\cal F}=\int d\mathbf{r}\left(\rho(-\frac{p^2}{2}+\frac{p^4}{4})+\frac{K}{2}(\partial_{\alpha}p_{\beta})(\partial_{\alpha}p_{\beta})\right). \label{eqn::Frank}
\end{equation}
where we use the Einstein's summation convention over Greek indices. The polynomial part favours the emergence of a finite polarity vector $p_\alpha$ of modulus $p=1$, while the second term penalizes energetically the formation of large gradients. 
The energy scale of the nematic elasticity is fixed by the parameter $\rho>0$, which in our 2d model have dimensions of stress. The balance between the two terms defines a characteristic scale of spatial variation of the polarity, the so-called nematic correlation length $L_{c}\equiv\sqrt{K/\rho}$. The conjugated field of the polarity, the so-called molecular field, is given by 
$h_\alpha=-{\cal \delta F}/{\cal \delta}p_\alpha$.

Cellular traction forces are transmitted across the epithelial monolayer, through cell-cell adhesion proteins
leading to complex self-organised collective patterns reflected in the motion of cells, \cite{Serra-Picamal2012a, Vedula2012, Deforet2014}. We assume a coarse-grained point of view at scales larger than the cell size, and propose a continuum description of the system. Building on previous models \cite{mercader2017effective}, we assume that at sufficiently long time scales, the medium can be described as a viscous fluid. This is consistent with the direct observation of the fluid-like relative motion of cells, and assumes that all the processes that control the elastic properties of the medium (excluding the nematic elasticity) relax in much shorter time scales. The kinetics of the cell-cell adhesion, for instance, has a turnover time scale that is in the range of $\sim 10$~min \cite{Lambert2007}, while the time scales of observation for the phenomena here studied are at least one order of magnitude larger. Additional  fluidization mechanisms have been reported that result from cell division \cite{Ranft2010}. While this effect will coexist, the rate of division is not very significant in the experiments that we are modelling \cite{Serra-Picamal2012a}, and we will thus neglect the possible active stresses generated by cell division. 

Since the cell monolayer is a quasi-two dimensional system, we will assume an effective 2d description, extending the approach of \cite{mercader2017effective}. We thus take the hydrodynamic equations describing a $2d$ (compressible) active polar fluid, with nematic elasticity, that are consistent with symmetries and include active and passive contact forces with the substrate. Our model is completely specified by the set of equations 
%
%
\begin{eqnarray}
&& \partial_{\beta}\sigma_{\alpha\beta}=\xi v_{\alpha}-T_{0}p_{\alpha},\label{eqn::forcebalance}
\\&&\sigma_{\alpha\beta}=2\eta v_{\alpha\beta}-\zeta(p_{\alpha}p_{\beta}-\frac{p^2}{2}\delta_{\alpha\beta})\nonumber
\\&&+\frac{(\nu+1)}{2}p_{\alpha}h_{\beta}
+\frac{(\nu-1)}{2}p_{\beta}h_{\alpha}-\frac{\nu}{2}p_{\gamma}h_{\gamma}\delta_{\alpha\beta},\label{eqn::totalstressexpression}
\\&& \partial_{t}p_{\alpha}+v_{\gamma}\partial_{\gamma}p_{\alpha}+\omega_{\alpha\beta}p_{\beta}=\frac{1}{\gamma_{1}}h_{\alpha}-\nu v_{\alpha\beta}p_{\beta},\label{eqn::polarity}
\\&& h_\alpha = \rho(1-p^2)p_\alpha + K\nabla^2 p_\alpha. \label{eqn::molecularfield}
\end{eqnarray}
where $\sigma_{\alpha \beta}$ is the traceless stress tensor, and $v_{\alpha\beta}$ and $\omega_{\alpha\beta}$ are the traceless symmetric and antisymmetric components of the velocity gradient tensor, respectively.

Eq.~(\ref{eqn::forcebalance}) expresses the force balance in the absence of inertia, with the total external force in the rhs, including a passive (friction) and active (traction) contributions. 
%
%
Eq.~(\ref{eqn::totalstressexpression}) is the constitutive equation for the total stress of an active polar liquid. It is assumed that in the time scales of observation, elastic effects can be neglected. The first term in the rhs, accounts for viscous stresses with $\eta$ being the shear viscosity. The second term accounts for active stresses, with the activity parameter $\zeta<0$ for contractile stress, which is the relevant case for epithelial monolayers. The rest of terms are the usual ones describing nematic elasticity \cite{de1993physics}. Eq.~(\ref{eqn::polarity}) describes the dynamics of the polarity field. The lhs is the total co-moving co-rotating derivative, and the rhs describes the rotational relaxation of the polarity, being $\gamma_1$ the rotational viscosity. The last term, which couples the polarity and the flow is the so-called flow-alignment contribution, and by virtue of the Onsager reciprocity relations, must be characterized by the same coefficient $\nu$ appearing in Eq.~(\ref{eqn::totalstressexpression}). 
Eq.~(\ref{eqn::molecularfield}) specifies the molecular field $\mathbf{h}$ in terms of the polarity consistent with Eq.~(\ref{eqn::Frank}). 

In addition to the active contractility, we have introduced an additional active term that accounts for traction forces exerted on the substrate. 
The proposed form of this external force has been used before in the context of cell monolayers \cite{Lee2011,Lee2011a,mercader2017effective}, as well as for bacterial suspensions \cite{Aranson2007}. The form is limited by symmetry considerations \cite{Julicher2009a} and can be derived from microscopic models with linker kinetics \cite{oriola2017fluid}. The form of the friction force, with an effective friction coefficient $\xi$ is standard \cite{Schwarz2013}. The active traction must be related to the consumption of ATP, as the active contractility term, but the two parameters $T_0$ and $\zeta$ originate at different mechanisms and can be considered as independent parameters. $T_{0}>0$ defines the scale of the cellular traction forces.

An effective 2d model such as the one proposed can in principle be derived from a 3d thin layer using the lubrication approximation and averaging over the monolayer thickness, as described in \cite{mercader2017effective} for a 1d reduction. The 3d incompressibility of the fluid then allows to eliminate the pressure from the description. In the reduced description, however, the fluid is compressible, and for a reduction to 2d, the constitutive equation for the trace of the stress tensor $\sigma$ must be specified. For simplicity we assume 
\begin{equation}
\sigma=\tilde{\eta}v_{\gamma\gamma}-\frac{\tilde{\zeta}}{2} p_{\gamma}p_{\gamma}+\frac{\tilde{\nu}}{2}p_{\gamma}h_{\gamma}=0,\label{sigma}
\end{equation}
with the three coefficients $\tilde{\eta}$, $\tilde{\zeta}$, $\tilde{\nu}$ each being zero. Strictly speaking, this choice 
gives up 3d incompressibility, since these parameters are not free within that condition. However we may still omit pressure gradients in the force balance Eq.~(\ref{eqn::forcebalance}) by assuming that the 2d effective fluid has a large compressibility, which accounts for the fact that an in-plane compression offers no significant resistance because it can be accommodated by an expansion in the third dimension. 

The parameters of the problem can be grouped in different useful combinations. 
In addition to the nematic correlation length $L_c \equiv \sqrt{K/\rho} $, we introduce two friction lengths defined by 
$L_\eta \equiv \sqrt{\eta/\xi}$ and $L_\gamma \equiv \sqrt{\rho\gamma/\xi}$ (with $\gamma \equiv \gamma_1/\rho$).
We use the nematic stress scale $\rho$ to define a dimensionless contractility $\bar{\zeta}=\zeta/\rho$ and a dimensionless traction $\bar{T}_0=T_0 L_\gamma / \rho$. In addition to $\nu$, $\bar{\zeta}$ and $\bar{T}_0$, the model contains the three length scales $L_c$, $L_\eta$ and $L_\gamma$ and one time scale $\gamma$. Using two of them to scale length and time, we are left with a set of five independent dimensionless parameters.
%

\section{Linear instability of a homogeneously polarized state}

\subsection{Linear stability analysis}

The set of equations (\ref{eqn::forcebalance}-\ref{eqn::molecularfield}) have trivial homogeneous solution with a uniform polarity field and a uniform velocity $V$ in the direction of the polarity field, with $V=T_0/\xi$. The direction of this fields has the continuous degeneracy of the rotational invariance of the problem, which is spontaneously broken. Without lack of generality, we chose the polarity and the velocity to be oriented in the $x$-direction so that $p_x=1$, $p_y=0$, $v_x=V$, $v_y=0$ with the component of the stress $\sigma_{xx}=-\zeta /2$. In this section we address the linear stability analysis of this base state, that is, we will obtain the growth rate $\omega({\bf q})$ of sinusoidal perturbations of the form $\exp{(\omega(q) t + i {\bf q}\cdot {\bf r})}$ under the linearized dynamics around the base state. We distinguish to two types of modes, transverse and longitudinal, which designate perturbations of the polarity with $\bf{q}$ parallel to the $x$-direction that are perpendicular ($\delta p_x=0$) and parallel ($\delta p_y=0$) respectively, or equivalently, that modify the direction and the modulus, respectively. At linear level these two types of modes are decoupled, so we find two branches for the linear growth rate  $\omega_L$ and $\omega_T$, which 
can be written as
\begin{equation}
\omega^{L,T}({\bf q}) = \Omega \left(\lambda \pm \sqrt{D} \right).
\label{eq::omegatot}
\end{equation}
From the anisotropy of the problem, this growth rate depends on the modulus $q^2 \equiv q_x^2+q_x^2$ and on terms of the form $q^n \cos{n\theta}$ where $\theta$ is defined by 
$\cos{\theta}=\hat{\bf q}\cdot \hat{\bf x}$. To make the notation more compact we introduce the complex wave vector $Q\equiv q_x+iq_y$, such that
Re$[Q^n]=q^n\cos{n\theta}$. Equivalently, we have Re$[Q^2]=q_x^2-q_y^2$, Re$[Q^3]=q_x(q_x^2-3q_y^2)$, 
and Re$[Q^4]=q_x^4 - 6 q_x^2 q_y^2 + q_y^4$.
We can now split the contributions to the dimensionless growth rate $\omega /\Omega$ according to passive, 
contractility and traction terms, such that we have 
\begin{eqnarray}
\lambda&=&\lambda_p+{\rm Re}[\lambda_\zeta]+i {\rm Re}[\lambda_{T_0}] \\
D&=&{\rm Re}[D_p+D_\zeta+D^R_{T_0}] + i{\rm Re}[D^I_{T_0}].
\end{eqnarray}
If we adopt the compact notation $q^2_c \equiv L^2_c q^2$, $q^2_\eta \equiv L^2_\eta q^2$ and $q^2_\gamma \equiv L^2_\gamma q^2$, and $Q_\gamma \equiv L_\gamma Q$, we have
\begin{eqnarray}
\Omega^{-1}&=&4\gamma(1+q_\eta^2) \nonumber \\
\lambda_p &=&\nu q_{c}^2 Q_{\gamma}^2 -\frac{1}{\gamma\Omega}(1+q_c^2) -\frac{1}{2}q_\gamma^2 (2\nu^2 + q_c^2(1+2\nu^2)) \nonumber \\
\lambda_\zeta &=& \bar{\zeta}(Q_\gamma^2-2\nu q_\gamma^2) \nonumber \\
\lambda_{T_0}&=&-\bar{T}_0 Q_\gamma (4q_\eta^2+2\nu+3) \nonumber
\end{eqnarray}
and 
%
%
%
\begin{eqnarray}
D_p &=& \frac{1}{\gamma^2\Omega^2} + \frac{1}{\gamma\Omega} q^2_\gamma (2\nu^2-q_c^2)  - \nu^2 q_c^2 Q_\gamma^4            \nonumber \\
&+& \frac{1}{4}q_\gamma^4 \left( 4\nu^4 + q_c^4 (1+ 4\nu^2 ) \right) 
\nonumber \\
&-& \nu q_c^2 Q_\gamma^2 \left( q_\gamma^2 (q_c^2 - 2\nu^2 ) - 8(1+q_{\eta}^2) \right)   \nonumber \\
D_\zeta &=&  \bar{\zeta} \nu \left[ 2 q_c^2 q_\gamma^4 - Q_\gamma^2 
\left( q_\gamma^2 (q_c^2 - 2\nu^2 ) - 8(1+q_{\eta}^2) \right) \right] \nonumber \\
&+&\frac{1}{2} \bar{\zeta}^2 (q_\gamma^4+Q_\gamma^4) \nonumber \\
%
D^R_{T_0}&=& \frac{1}{2}\bar{T}_0^2 \left[ (q_\gamma^2-Q_\gamma^2) (2\nu+1)^2 -2q_\gamma^2 \right]   \nonumber\\
D^I_{T_0}&=& \bar{T}_0 Q_\gamma \left[\left( 8(1+q_\eta^2) - q_\gamma^2(\bar{\zeta}-q_c^2 )(2\nu-1) 
\right) \right. \nonumber \\
&+& \left. Q_\gamma^2 \left( \bar{\zeta} (2\nu+1)+2\nu^2 \right) \right] \nonumber
\end{eqnarray}
%
%

The above expressions contain a wealth of physical information about the dynamics of the system already at linear level. The spontaneously broken isotropy is captured by the terms containing $Q$. 
The prefactor $\Omega(q)$ in Eq.~(\ref{eq::omegatot}) is a Lorentzian propagator that expresses the nonlocal character of the dynamics (see also discussion of section III.C). In addition, the contributions contained in $\sqrt{D}$ are also nonlocal, since they involve all orders in $q$.
Note also that, because of the symmetries of the problem, the linear growth rate satisfies 
 $\omega(-{\bf q})=\omega^*({\bf q})$, where the asterisk denotes complex conjugate, $\omega (-q_y) = \omega (q_y)$ for any $q_x$, and, for $q_x=0$ we have Im$[\omega] = 0$. From a theoretical point of view, and for further reference it is interesting to write explicitly the growth rates in the long wavelength limit, as an expansion up to order $q^2$. These read
\begin{eqnarray}
\gamma \omega^L &=& -2 + q_\gamma^2 \left( \frac{\nu\bar{\zeta}}{2} + \frac{\nu^2}{2} +\frac{L_c^2}{L_\gamma^2}          +\frac{\nu(\nu+1)}{8}\bar{T}_0^2 \sin^2{\theta} \right)
\nonumber \\
&\;& - i q_\gamma \frac{\nu+2}{2}\bar{T}_0  \cos{\theta} + {\cal O}(q^4), \\
\gamma \omega^T &=& - q_\gamma^2 \left(\frac{ \bar{\zeta}(\nu-\cos{2\theta})}{2} + \frac{L_c^2}{L_\gamma^2}         
-\frac{\nu(\nu+1)}{8} \bar{T}_0^2 \sin^2{\theta} \right)
\nonumber \\
&\;& - i q_\gamma\frac{\nu+1}{2}\bar{T}_0  \cos{\theta} + {\cal O}(q^4).
\end{eqnarray} 
The hydrodynamic limit $q\rightarrow 0$ is fundamentally different for the two types of modes. The transverse zero-mode is marginal, i.e. $\omega^T(q=0)=0$, because of the rotational invariance of the problem, since this mode accounts for an infinitesimal homogeneous rotation of the base state, which has a continuous degeneracy. This soft mode associated to the rotational symmetry will play an important role in the discussion of possible routes to chaos in this problem in Section.~\ref{sec:nonlinearwaves}. 
By contrast, the longitudinal modes relax in a time of order one in this limit, and hence they are not hydrodynamic. The separation of time scales between the relaxation of the modulus of the polarity vector and its direction is often invoked to justify an adiabatic elimination of the dynamics of the modulus. In our case, however, the region of interest for the experiments is not in the hydrodynamic limit and longitudinal modes will be essential to interpret the observations. 

\subsection{Bifurcation into travelling waves}

We find that, in general, both $\lambda$ and $D$ are complex numbers, so the dispersion relations $\omega^{L,T}$ contain both a real and an imaginary part. The condition Re$[\omega]=0$ defines an instability boundary.
If Im$[\omega] \neq 0$ when Re$[\omega]=0$, then the instability is oscillatory (i.e. a Hopf bifurcation) giving rise to travelling waves. 

From the explicit expression of the growth rates, we see that the passive contribution to Re$[\omega]$ is always negative, and the base state is stable. However, if the system is sufficiently active, there may exist modes with Re$[\omega]>0$ and thus an instability sets in. Hereinafter we will restrict to the case $\nu>0$ which is the one relevant for the discussion of spreading epithelial monolayers. Then, for longitudinal modes, the instability can only happen for contractile active stress (i.e. $\zeta<0$) and for a band of modes with finite ${\bf q}$ around the critical wave number ${\bf q_o}$ for which Re$[\omega^L({\bf q_o})]=0$,  excluding the neighborhood of the mode ${\bf q}=0$, which is always stable. An unstable band thus exists for $|\bar{\zeta} | > 
|\bar{\zeta}^L_o|$ where the threshold value of contractility for the onset of the longitudinal instability is given by
\begin{equation}
|\bar{\zeta}^L_o|=\frac{2}{\nu}\frac{L_c^2}{L_\gamma^2} \left( 1+ \frac{L_\eta}{L_c}\sqrt{2+\frac{\nu^2}{2}\frac{L_\gamma^2}{L_\eta^2}   }      \right)^2. \label{eq:threshold}
\end{equation}
In this case, the onset of instability occurs at a finite $q_o$. This situation is usually referred as an oscillatory periodic instability (see \cite{cross1993pattern}). For the $1d$ case (i.e. $q_y=0$) the critical mode where the instability sets in is
\begin{equation}
q_o^2=\frac{1}{L_c L_\eta} \sqrt{\frac{2}{1+\frac{\nu^2L_\gamma^2}{4L_\eta^2} }}.
\end{equation}
Since Re$[\omega^L({\bf q})]$ is arbitrarily small close to threshold, while the frequency Im$[\omega^L({\bf q})]$ remains finite, slightly above threshold a localized perturbation is advected faster than it grows, so the perturbed fields eventually relax to the unperturbed values, a situation that is referred to as convective instability. 
Only over a finite distance from threshold, the system is said to be absolutely unstable, that is, that a localized perturbation at a given location does grow in amplitude at that location.  Note that for the longitudinal modes, the instability requires a finite value of the flow alignment coefficient $\nu$. Large values of $\nu$, and small values of friction favour the instability. 

For the transverse modes, the instability is controlled by both active parameters. The condition for the instability then reads
\begin{equation}
\bar{\zeta} - \frac{\nu}{4} \bar{T}_0^2 <  -\frac{2}{\nu+1} \frac{L_c^2}{L_\gamma^2}. \label{eq:onset_trans}
\end{equation}
Remarkably, in this case, the instability may be driven solely by traction forces and may occur even for extensile activity (i.e. $\zeta>0$). The transverse instability is a long-wave length one, that is with the critical wave vector ${\bf q_o}=0$. At a finite value above threshold, then Re$[\omega^T]$ is peaked at finite $q$ but remains marginal at $q=0$, as imposed by the rotational invariance of the problem. Note also that the transverse instability sets in before the longitudinal one as we increase $|\zeta |$, since the condition (\ref{eq:onset_trans}) is satisfied for $\bar{\zeta}^L_o<\bar{\zeta}$, even if $\bar{T}_0=0$.

We remark that, for any $q_x \neq 0$, both longitudinal and transverse modes have a non-zero imaginary part of $\omega^{L,T}$ provided that $T_0 \neq 0$. Therefore, the presence of traction forces is directly associated 
to oscillatory behavior. 
In addition, the sign of Im$[\omega]$ is always opposite to that of $q_x$, implying that there will be waves travelling only in the direction of the polarity ${\bf p}$ (from negative to positive $x$ in our case), reflecting the anisotropy of the base state. Similarly, the fact that for $q_x=0$ we have Im$[\omega] = 0$ implies that waves travelling along the $y$-axis are not possible. 

The linear instability criterion does not exclude that finite-amplitude nonlinear waves or other type of solutions can exist even in the linearly stable regions, in particular near the linear stability boundary. This nonlinear instability of the base state, is characteristic of subcritical bifurcations and implies that sufficiently large finite-amplitude perturbations of the base state may grow even if smaller amplitudes do decay. In the analysis of the following sections we will determine exactly the boundaries where the bifurcation in our model is subcritical.

\begin{figure}[t] 
   \includegraphics[width=8.6cm]{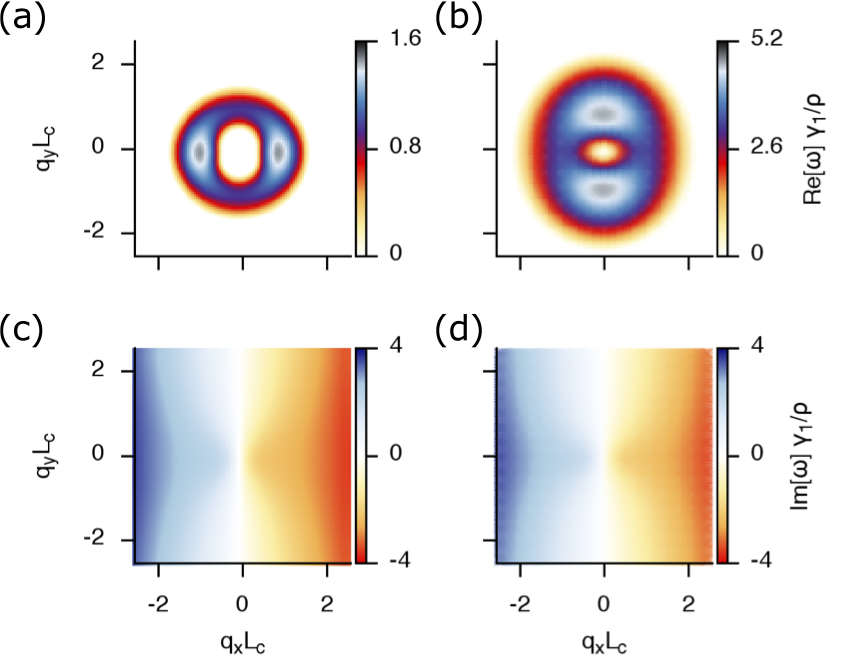}
   \caption{Linear growth rate of small perturbation of a homogeneously polarized state. The upper (lower) row displays the positive real (imaginary) part of the growth rate in the $(q_{x},q_{y})$ plane. The left (right) column represents the growth rate of the longitudinal (transverse) modes, with color-coded amplitude. The values of parameters are: $\eta=10^6$, $\xi=100$, $T_{0} =10$, $L_{c} =50$, $\gamma=600$, $\nu=10$, $\rho=10$ and $\zeta=-2000$, in units of Pa, $\mu$m and s.} 
    \label{Fig::1}
\end{figure}

In Fig.~(\ref{Fig::1}) we plot the real and imaginary parts of both branches $\omega^{L,T}$, for parameter values relevant to MDCK cells \cite{Serra-Picamal2012a, Vedula2012, Deforet2014}. We observe that the transverse instability appears first and the growth rate is peaked at finite $q_y$ and $q_x=0$. By contrast, the longitudinal instability peaks at finite $q_x$ and $q_y=0$. 

The explicit expressions for the growth rate on the $x,y$-axes take a simpler form. 
For $q_y=0$ and writing $q\equiv q_x$ and $\omega_x (q) \equiv \omega(q_x)$, we have
\begin{eqnarray}
\gamma \omega_x^L &=& -(2+q_c^2) \left( 1+\frac{\nu^2 q_\gamma^2}{4(1+q_\eta^2)}   \right) - 
\bar{\zeta}\frac{\nu q_\gamma^2}{2(1+q_\eta^2)}  \nonumber \\
&\;&  -i q_\gamma \bar{T}_0 \left(1+ \frac{\nu}{2(1+q_\eta^2)}  \right)
  \label{eq:long_x} \\
\gamma \omega_x^T &=& -q_c^2 \left( 1+\frac{(\nu-1)^2 q_\gamma^2}{4(1+q_\eta^2)}   \right) -
\bar{\zeta}\frac{(\nu-1) q_\gamma^2}{2(1+q_\eta^2)} \nonumber \\
&\;& - i q_\gamma \bar{T}_0 \left(1+ \frac{\nu-1}{2(1+q_\eta^2)}  \right)
  \label{eq:trans_x}
\end{eqnarray}
According to these results, the characteristic frequency scale of the oscillations $w_o=|{\rm Im}[\omega(q_o)]|$ is given by $\omega_o \sim T_0 q_o / \xi$. Therefore, the origin of the slow time scale of oscillations is not intrinsic of the material properties but depends on the interaction with the substrate, which fixes the friction and traction forces. In Section~\ref{sec:discussionandconclusions} we will see that the observed waves in experiments can indeed be interpreted as those resulting from this instability.

The phase velocity of a wave with wave-vector $q_x$ for the two types of modes are given by $v^{L,T}= | {\rm Im}[\omega^{L,T}_x]|/q_x$. Using that the flow velocity of the base state is $V=T_0/\xi$, we find that 
\begin{eqnarray}
\frac{v^L}{V}=1+\frac{\nu}{2+q_\eta^2}, \label{eq:vL}\\
\frac{vT}{V}=1+\frac{\nu-1}{2+q_\eta^2}.\label{eq:vT}
\end{eqnarray}
This result allows us to infer a value of the parameter $\nu$, which is usually unknown for tissues, just comparing the wave velocity and the front velocity in the experiments of Ref. \cite{Serra-Picamal2012a} (see Section~\ref{sec:discussionandconclusions}). 

In the $y$-axis, $q_x=0$, we have Im$[\omega]=0$. The presence of traction modifies quantitatively the growth rate
but this is qualitatively similar to that the case $T_0=0$, which reads, with $q\equiv q_y$,
\begin{eqnarray}
\gamma \omega_y^L &=& -(2+q_c^2) \left( 1+\frac{\nu^2 q_\gamma^2}{4(1+q_\eta^2)}   \right)  
-\frac{\nu\bar{\zeta} q_\gamma^2}{2(1+q_\eta^2)}  
  \label{eq:long_y} \\
\gamma \omega_y^T &=& -q_c^2 \left( 1+\frac{(\nu+1)^2 q_\gamma^2}{4(1+q_\eta^2)}   \right) 
-\frac{(\nu+1)\bar{\zeta} q_\gamma^2}{2(1+q_\eta^2)}. 
  \label{eq:trans_y}
\end{eqnarray}  
 
\subsection{Physical origin of waves and phase lag}

In most of the subsequent analysis, we will pursue the study of the $1d$-case corresponding to longitudinal modes with $q_y=0$ (Eq.~(\ref{eq:long_x})). This is the simplest case and at the same time 
the most interesting to gain insights into the physical mechanism behind the waves, to analyze in depth the nonlinear dynamics of the problem, and to compare with experiments.

In this case the model equations reduce to
\begin{eqnarray}
&& \partial_{x}\sigma_{xx}=\xi v_{x}-T_{0}p_{x},\label{eqn::forcebalance1}
\\&&\sigma_{xx}=\eta \partial_x v_{x}-\frac{1}{2}\zeta p_x^2 -\frac{\nu}{2}p_x h_x,
\label{eqn::totalstressexpression1}
\\&& \partial_{t}p_{x}+v_{x}\partial_{x}p_{x} =\frac{1}{\rho \gamma}h_{x}-\frac{\nu}{2} p_x\partial_x v_{x},\label{eqn::polarity1}
\\&& h_x = \rho(1-p_x^2)p_x + K\partial^2_x p_x  \label{eqn::molecularfield1}.
\end{eqnarray}

We remark that the emergence of elastic-like waves appears naturally in our model for a purely viscous constitutive equation as long as active traction is present, without invoking any additional time scale related to an extra coupling to internal variables, as is usually required in models based on an elastic constitutive equations~\cite{Serra-Picamal2012a,notbohm2016cellular,tlili2016waves}. For the sake of discussion, let us consider the simple case of $\nu=0$ and $\zeta =0$. 
At linear level, the evolution of a small perturbation $\delta p$ is coupled to traction through the advective term, 
\begin{equation}
\partial_t \delta p + V \partial_x \delta p = \frac{1}{\gamma} \left(  -2 \delta p + L_c^2 \partial_x^2 \delta p  \right),
\label{eqn::linearp}
\end{equation}
with $V=T_0/\xi$, implying that a perturbation of the polarity proportional to $\exp{(\omega(q) t + i {\bf q}\cdot {\bf r})}$ will decay but at the same time be advected with the velocity $V=T_0/\xi$. We thus have a dispersion relation with
Im$[\omega] = - V q$. 
This is reminiscent of that of elastic waves (i.e. Im$[\omega] = \pm V |q|$), but with the fundamental difference that we get only one propagation direction, due to the fact that rotational symmetry (or parity in the $1d$ case) are explicitly broken by the base state. The key observation is that we can close an equation for $p$ that is of first order in time derivatives. This is in contrast to the case where the medium is assumed elastic. As pointed out in Refs.~\cite{wavesMarchetti2015,notbohm2016cellular}, in that case an effective inertia must be invoked to obtain elastic waves. In Ref.~\cite{wavesMarchetti2015}, this is achieved by introducing an additional coupling to a slow variable, such that, at linear level, a wave equation (i.e. with second order time derivatives) is obtained for the strain field. Note that the resulting elastic waves are apolar, that is, insensitive to the sign of $p$.
In a purely viscous medium, however, we do obtain propagating waves through the advective coupling, as long as the force balance equation includes a finite traction $T_0$, and these waves are polar. The other observables can be directly related to $\delta p$ obtained from Eq.~(\ref{eqn::linearp}). Note that the combination of the force balance equation and the viscous constitutive equation imply that the relationship between $\delta p$ and both $\delta v$ and $\delta \sigma$ are nonlocal 
with
\begin{equation}
\delta v (x) =  \frac{V}{2 L_\eta}\int e^{-|x-x'|/L_{\eta}} \; \delta p(x')dx' .
\label{eqn::nonlocal}
\end{equation}
and $\delta \sigma (x)= \eta \partial_x \delta v(x)$.


The scenario for the emergence of waves in our model is thus fundamentally different from that provided by an elastic medium with an effective inertia. Our scenario is similar, if we reintroduce non-zero values of $\nu$ and $\zeta$. Then the closed equation for $\delta p$ becomes nonlocal but still of first order in time derivatives 
\footnote{Note that the linear equation for the fluctuations of the strain field in the model of Ref.~\cite{wavesMarchetti2015,notbohm2016cellular} is of second order in time and includes viscoelasticity and dispersivity, but is still local in space}.
The physical picture in thus essentially the same with two important additional features. First, the presence of contractility can reverse the damping and generate the growth of the wave amplitude, thus allowing for sustained (nonlinear) waves. Second, the presence of either one of the two parameters is sufficient to introduce an elastic phase shift between stress and strain rate. The presence of such phase lag in the experiments of Ref.~\cite{Serra-Picamal2012a,notbohm2016cellular} has been widely interpreted as a signature of an elastic constitutive equation of the medium. Here we show that this inference is not really justified, since it is also possible to have the same phase lag in a purely viscous medium, provided that active traction is present ($T_0 \neq 0$), together with at least one of the two parameters $\nu$ or $\zeta$ being nonzero. 

Solving the linearized equations Eqs.~(\ref{eqn::totalstressexpression1}-\ref{eqn::molecularfield1}) in Fourier space, the relative phase between stress and strain rate can be found exactly and reads 
\begin{equation}
\frac{\delta \hat{\sigma}_{xx}}{ i q \eta \delta \hat{v}_{x}}=
- \frac{ \bar{\zeta} + 2 \nu (1+q_{c}^2/2) - i q_{\gamma} \bar{T}_{0}L_\eta^2/L_\gamma^2}{q_{\eta}^2 ( \bar{\zeta} + 2  \nu  (1+q_{c}^2/2) + i \bar{T}_0/q_{\gamma})}.\label{eqn::generalfase}
\end{equation}
The rhs of Eq.~(\ref{eqn::generalfase}) being real would be characteristic of purely viscous medium. The presence of traction forces $T_0$, however, introduces an imaginary part that produces a phase lag that would be characteristic of elastic behavior. Note that the presence of nematic elasticity in the constitutive equation (i.e. the term proportional to $\nu$ in Eq.(\ref{eqn::totalstressexpression1})) would not be sufficient to introduce the elastic-like phase shift in the absence of traction.

In the particular case of $2 \nu (1+q_{c}^2/2) \ll |\bar{\zeta}|$ and $|\bar{\zeta}|q_{\gamma} \ll \bar{T}_0$, Eq. 
(\ref{eqn::generalfase}) takes the simple form 
\begin{equation}
\delta \hat{\sigma}_{xx}(q) \approx \left( \eta + i\frac{\zeta \xi}{q T_0}  \right) iq \;\delta \hat{v}_x (q).
\end{equation}
The phase lag in the spatial dependence will produce a phase lag in the time oscillations of the two observables at any given location, provided that travelling waves are sustained, that is, for $|\zeta|>|\zeta_o^{L}|$, and consequently $q \sim q_o$. We thus obtain that, if $|\zeta_o^{L}| \gg \omega_o \eta=T_{o}q_{o}\eta/\xi$ the phase lag for the observed waves will typically be that of an elastic medium, even though the rheology of the system is that of a purely viscous material. 

\begin{figure}[t] 
   \includegraphics[width=8.6cm]{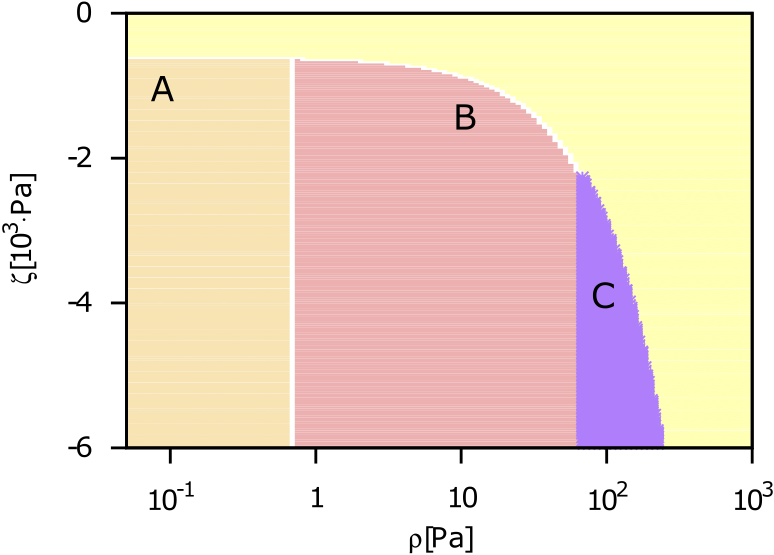}
   \caption{Phase-diagram of different nonlinear dynamics in the $\zeta$-$\rho$ plan. The yellow domain corresponds to the linearly stable region while the other domains correspond to nonlinear oscillatory  solutions displayed in Fig.~(\ref{Fig::3}). The boundary of the yellow domain is given by the longitudinal instability threshold $\zeta_{o}^{L}$. The other  borders are determined by the Complex-Ginzburg Landau Equation~(\ref{CGLE}). The parameters are: $\eta=10^6$, $\xi=100$, $T_{0}=10$, $L_{c}=50$, $\gamma=600$ and  $\nu=20$, in units of Pa, $\mu$m and s.}
    \label{Fig::2}
\end{figure}

\begin{figure}[t] 

   \includegraphics[width=8.6cm]{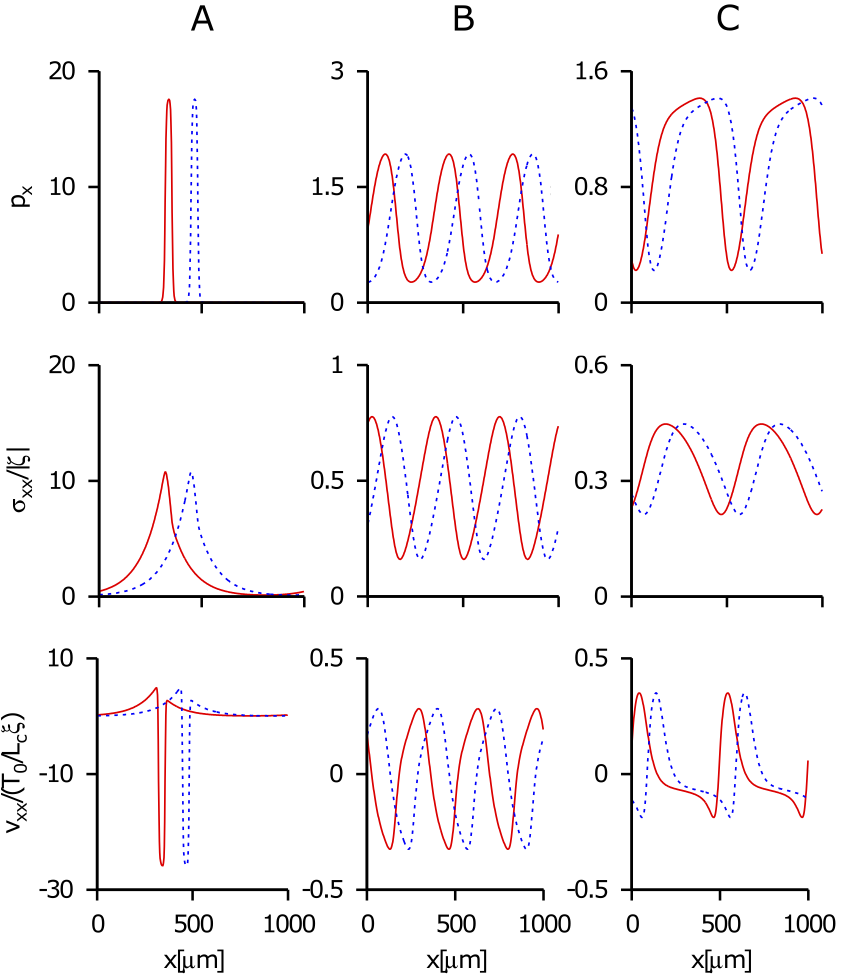}
   \caption{Representative steady oscillatory nonlinear profiles for the different domains of the phase diagram of  
   Fig.~(\ref{Fig::2}),  with the same set of parameters. 
   In the first column (A) $\rho=0.1$ and $\zeta=-1013$ in the second column (B) $10$ and $-1273$ and in the third column (C) $100$ and $-3398$, both coefficients in units of Pa. The solid red and dashed blue curve are spaced by a time lapse of $6$~min.} 
    \label{Fig::3}
\end{figure}

\section{Nonlinear waves}\label{sec:nonlinearwaves}

In the previous section we have analyzed the linearized dynamics around the homogeneously polarized state, and have identified broad regions of parameters where this state is unstable. The amplitude of unstable modes will grow until saturation by nonlinearities occurs. As we will show, the observed waves in a variety of experiments can be identified with such nonlinear waves. In this section we pursue the numerical and analytical study of such nonlinear waves. 
We focus most of the nonlinear analysis on the case of $1d$ longitudinal waves, and address only briefly more general situations. 

\subsection{Numerical analysis of longitudinal waves}

The numerical exploration of the five-dimensional parameter space of our problem is obviously prohibitive, so in a first numerical exploration of the deeply nonlinear regime we will construct a phase diagram $\zeta-\rho$, relating the main parameter driving the instability and the stress scale of the nematic elasticity, a phenomenological parameter which is rather elusive for cell tissues. With the rest of parameters fixed, we have that $\rho \propto L_\gamma^2$. 

We have integrated numerically the full nonlinear dynamics of the $1d$ model for longitudinal modes using a semi-implicit algorithm (Eqs.~(\ref{eqn::forcebalance1}-\ref{eqn::molecularfield1})). For parameter values in the range relevant to experiments, we have identified different types of solutions resulting in the phase-diagram plotted in Fig.~(\ref{Fig::2}). Beyond the threshold of the longitudinal oscillatory instability (i.e. $|\zeta| > |\zeta_o^{L}|$) we find three classes of nonlinear solutions. In domain $A$, we observe that the polarity field, after a random perturbation, develops a transient array of equally spaced localized pulses, which after some time coalesce giving rise to the formation of an isolated pulse of polarization propagating through a nonpolarized medium, as shown in  Fig.~(\ref{Fig::3}). The maximal value of the polarity field is of the order of $\sim\sqrt{|\zeta|/\rho\nu}$ and the transition from the homogeneously polarized state is discontinuous, that is, implying that the propagating solutions appear at a finite amplitude right on threshold (subcritical bifurcation). In domains $B$ and $C$, the physical observables exhibit a nonlinear travelling periodic pattern. In both cases, the travelling wave speed is of the order of $ T_{0}/\xi$ and the spatial periodicity is of the order of $1/q_{o}$. The transition from the homogeneous polarized state is continuous (supercritical) for domain $B$ and discontinuous (subcritical) for domain $C$. The rest of the diagram, plotted in yellow, corresponds to the linearly stable region. 


\subsection{Complex Ginzburg-Landau equation} 


The above solutions correspond to regions arbitrarily far from the instability threshold and are illustrative of typical solutions that could be identified with experimental observations, where both wave trains and solitary fronts have been reported \cite{Serra-Picamal2012a, Vedula2012, Deforet2014}. However, it is obviously unpractical to explore numerically the different qualitative scenarios of nonlinear dynamics in a five-dimensional parameter space. Alternatively, one
may exploit the universality of the dynamics of any nonlinear system close to an instability, to develop 
a reduced description of the nonlinear dynamics of our model near threshold. Such a center-manifold 
projection does capture the essential nonlinear features of the problem \cite{cross1993pattern}. The resulting description depends only on symmetries and the nature of the bifurcation, and thus allows to classify the nonlinear dynamics regardless of the physical mechanisms responsible for the instability. This is a powerful approach that can be carried out analytically in the framework of a formal expansion on a small parameter $\epsilon$ defined by the normalized distance to threshold, and exploits the separation of length and time scales of the spatiotemporal variations of the bifurcating modes with respect to the original scales of the system.  
In this framework, the so-called amplitude equation appears as a solvability condition at lowest nontrivial order within that expansion. In our case $\epsilon \equiv 
(\zeta - \zeta_o^L)/\zeta_o^L$, which is positive (above threshold) when $\zeta < \zeta_o^L < 0$.
We define the complex envelope $A$ that describes the modulus and phase modulations of the  bifurcating mode in the form 
\begin{equation}
\delta p(x,t) \sim A(X,T)e^{i(q_o x - \omega_o t)}
\end{equation}
where $q_{o}$ is the critical wave vector and $\omega_{o}$ its corresponding frequency and the slow variables denoted by capital letters are defined as $T=\epsilon t$ and $X=\epsilon^{1/2}(x+V_gt)$, where $V_g$ is the group velocity of the envelope. The amplitude of the wave modulation of the base state is $\delta p(x,t) \sim \epsilon^{1/2}$. Since in our case, the symmetry $x\rightarrow -x$ is broken and waves only travel in one direction, we will have only the amplitude equation for the right-traveling wave. 
Accordingly, we can formulate the problem in a reference frame travelling with the group velocity. The normal form is then that of the so-called uniform oscillatory instability (i.e. $\omega_o\neq 0$, $q_o=0$) \cite{cross1993pattern}, even though we have $q_o \neq 0$. 
After proper rescaling, the normal form for the range above threshold 
is the so-called Complex Ginzburg-Landau equation CGLE \cite{cross1993pattern,aranson2002world},
\begin{equation}
\partial_T A =  A + (1+ib)\partial^2_X A - s (1+ic) |A|^2 A. \label{CGLE}
\end{equation}
The parameter in front of the cubic nonlinearity is $s=\pm 1$. If $s=1$, the equation corresponds to the case of the bifurcation being supercritical (continuous). 
For $s=-1$ the bifurcation is subcritical (discontinuous) and the equation must be supplemented by a quintic term.  

It is worth remarking that, once an additional parameter is eliminated by the condition of being at the instability threshold, the original five-parameter problem gets reduced to a two-parameter problem. 
Accordingly, since the phase diagram of Eq.~(\ref{CGLE}) is well-known \cite{chate1994spatiotemporal}, to classify all possible nonlinear scenarios (near threshold) it suffices to obtain the map of the physical parameters of the original problem into the parameters of the amplitude equation $c$ and $b$. This mapping always exists but it may be difficult to obtain in practice. We have performed this analysis following standard methods and with the help of the software Mathematica.
The extent to which the results obtained through this weakly nonlinear analysis apply to situations more deeply in the nonlinear regime is not guaranteed, 
but it is plausible to expect that the qualitative behavior will be similar. 

The crucial piece of information is thus the explicit map that relates $c$ and $b$ to the physical parameters, and the region of parameters for which $s=1$. In terms of the conveniently redefined set of dimensionless parameters 
$F \equiv \bar{T}_0 L_\gamma / L_\eta$, $G\equiv L_c/L_\eta$ and $H\equiv \nu^2 L_\gamma^2/ L_\eta^2$, 
in the Appendix A we derive the explicit maps
\begin{eqnarray}
b &=& b \;(\nu, F, G, H), \\
c &=& c \;(\nu, F, G, H), \\
s &=& s \;(\nu, F, G, H).
\end{eqnarray}
Note that the dimensionless contractility $\bar{\zeta}$ does not appear, because it has been eliminated by the additional constraint of being at the instability threshold $\bar{\zeta}=\bar{\zeta}^L_o(\nu,G,H)$.

From Eq.~(\ref{eq:threshold}) we thus have
\begin{equation}
|\bar{\zeta}^L_o|=\frac{2\nu}{H}\left(G+\sqrt{2+\frac{H}{2}}\right)^2.
\end{equation}
We then find
\begin{eqnarray}
b&=&\frac{F}{2}\frac{(4-3G\sqrt{8+2H})}{\sqrt{G}\sqrt[4]{8+2H} (8+G\sqrt{8+2H})^2}, \label{eq:b}\\
c&=&-\frac{f_{2}f_{4}-f_{1}f_{3}}{|f_{1}f_{2}+f_{3}f_{4}|}, \label{eq:c}\\
s&=&-{\rm Sign}\left(\frac{f_{1}f_{2}+f_{3}f_{4}}{f_{2}^2+f_{3}^2}\right), \label{eq:s}
\end{eqnarray}
where $f_i (\nu,F,G,H)$ are functions given explicitly in the Appendix A. The parameter $b$ has a much simpler expression because it depends solely on the linear part of the dynamics. The genuinely nontrivial part of the dynamics is contained in the parameters $c$ and $s$. The regions where the bifurcation is subcritical ($s=-1$) must be analyzed separately and will not be addressed here. 

\begin{figure}[t] 
   \includegraphics[width=8.6cm]{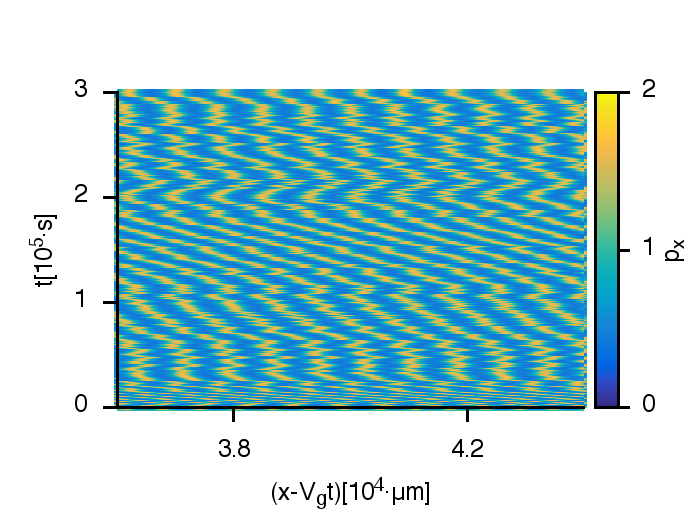}
   \caption{Phase turbulence. Numerical simulation of waves in the $1d$ model given by Eqs.~(\ref{eqn::forcebalance1}-\ref{eqn::molecularfield1}) for a region of the parameter space beyond the Benjamin-Feir instability line. The total size of the system is $8\cdot 10^4$, hence the kymograph is only representing $\sim 10\%$ of it. The x-axis is the position in a reference frame moving at the group velocity $V_{g}=1.41$ and the $y$-axis is time. $V_{g}$ has been estimated from the average of the phase velocity of polarity peaks. The colour bar labels the modulus of the director field $p_{x}$. The parameters are: $\eta=10^6$, $\xi=5$, $T_{0}=40$, $L_{c}=50$, $\bar{\gamma}_{1}=600$, $\nu_{1}=20$, $\zeta=-662.3$ and $\rho=10$, in units of Pa, $\mu$m and s.}
    \label{Fig::4}
\end{figure}

In the particular case of $T_0=0$, the bifurcation is of the stationary periodic type \cite{cross1993pattern}, with finite $q_o$ and $\omega_o=0$. Then the instability will lead to the formation of spatial patterns such as in Ref.~\cite{bois2011pattern} and the corresponding amplitude equation will be the so-called Real Ginzburg-Landau Equation, with $b=c=0$ \cite{cross1993pattern}. In this case, the dynamics of the amplitude equation is variational, that is, a Lyapunov functional ${\cal L}$ exists such that $\partial_T{A}=-\delta {\cal L} / \delta A^*$. Interestingly, this is no longer true if at least one of $b$ and $c$ are nonzero. In this case, the dynamics is said to be persistent and does not relax asymptotically to a specific pattern. This open the way to different forms of spatio-temporal chaos. The two-dimensional phase diagram of the $1d$ CGLE has been established in detail \cite{chate1994spatiotemporal,aranson2002world}, and is indeed extremely rich. Different complex dynamical regimes were identified such as the so-called phase turbulence, amplitude turbulence, spatio-temporal intermittency, and bistable chaos, in addition to regimes with more regular behavior. We may refer generically to the above classes of irregular persistent dynamics as weak turbulence. The boundaries of the different dynamical regimes are usually determined numerically. However, there is an exact boundary
that is relevant to our analysis, which locates the so-called Benjamin-Feir (BF) instability. This is a long wavelength  instability of the nonlinear travelling-wave solutions of the CGLE of the form $A_Q(X,T)=\sqrt{1-Q^2}\exp{(i(QX - T\Omega_Q}))$, with $\Omega_Q=c + (b-c) Q^2$. Such waves are unstable when the Newell criterion $1+bc < 0$ is satisfied. Beyond the BF line, one possibility is to 
have phase turbulence, where the wave phase changes in an irregular manner but preserving the winding number. In that regime, close to the BF line, the phase dynamics can be approximated by the Kuramoto-Sivashinsky equation \cite{cross1993pattern,chate1994spatiotemporal}. In other regions, crossing the BF line may lead to amplitude turbulence where the wave amplitude can reach zero values giving rise to non-conservation of the winding number. 
In the BF-stable side, however, one may also find regions with spatio-temporal intermittency, where patches of regular and chaotic behavior coexist. 
All these qualitative behaviors will be contained necessarily in our original model provided that the corresponding values of $c$ and $b$ can be reached by changing the model parameters. In Fig.~(\ref{Fig::4}) we show an example of phase turbulence for a situation in the BF-unstable region, obtained from numerical simulations of the original model in the appropriate parameter region. 

%

\begin{figure*}[t] 
   \includegraphics[width=12cm]{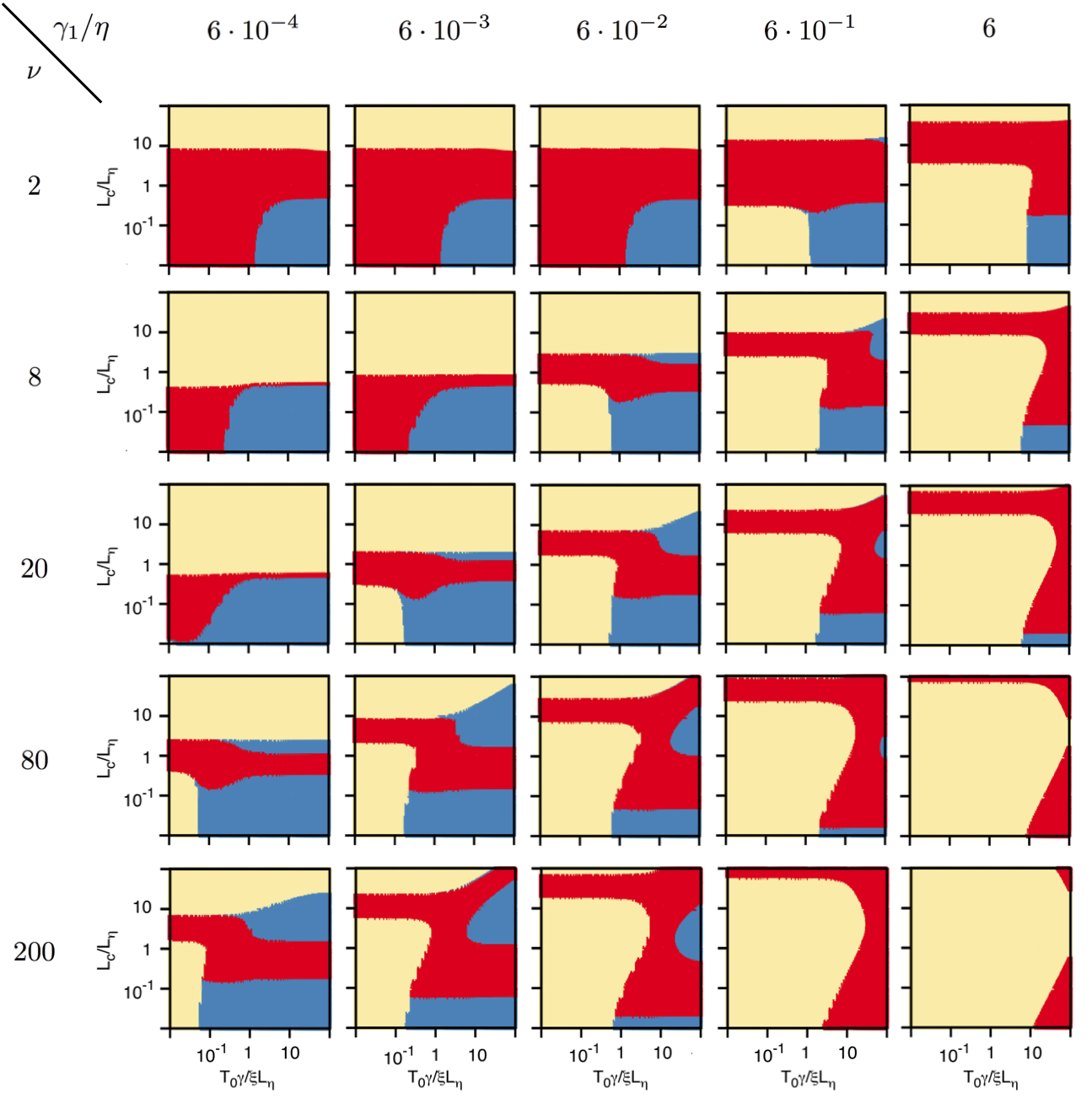}
   \caption{Exact subcritical-supercritical and Benjamin-Feir instability boundaries in a 4-dimensional plot of the model parameters, at onset of the primary instability of the problem. In the yellow region the primary instability is subcritical (discontinuous). In the red region the primary instability is supercritical and waves are BF-stable ($1+bc >0$). In the blue region, the primary instability is supercritical and waves are BF-unstable ($1+bc <0$). Each figure represents a cross-section of the 4-dimensional parameter space in the plane given by $T_{o}\gamma/\xi L_{\eta}$ and $L_{c}/L_{\eta}$, which are varied $4$ decades each. 
The dimensionless parameters $\gamma \rho/\eta$ and $\nu $ vary as indicated.
}
    \label{Fig::5}
\end{figure*}

To illustrate the complexity of the phase diagram in the four-dimensional space of the model projected in the region close to the instability threshold, we plot it in Fig.~(\ref{Fig::5}) for some ranges of the physical parameters. For simplicity we only distinguish the subcritical region, and supercritical region, which is split in the BF-stable and BF-unstable regions. In the latter, one may find phase turbulence, amplitude turbulence and bistable chaos. In the BF-stable region,
traveling-waves are linearly stable to long wavelength modulations, but spatio-temporal intermittency can also be found. 

We remark that the explicit knowledge of the exact maps Eqs.~(\ref{eq:b}-\ref{eq:s}) in a problem with so many parameters and with such a rich variety of complex nonlinear behavior, is extremely valuable. Indeed, 
for any set of physical parameters of the model, one can immediately find out the expected nonlinear behavior by checking the corresponding point in the known $b-c$ phase diagram, which is universal, and determined once for all.

%

\subsection{Transverse modes and soft-mode turbulence}

The weakly nonlinear analysis of the previous section is suitable for the case where $q_o \neq 0$, implying that near threshold, there is a narrow band of nearly marginal modes that excludes $q=0$. 
For transverse modes, however,
$q_o=0$, and above threshold the band of unstable modes extends all the way to $q=0$, which remains marginal because of rotational invariance. A 1d amplitude equation for the transverse mode along the $y$ axis can also be derived, now for a real amplitude field. We do not address this case here but, as discussed in \cite{cross1993pattern}, we can remark that this scenario also includes the possibility of phase turbulence, in the form of the Kuramoto-Sivashinsky equation. 


In general, for a 2d case the longitudinal and transverse modes will be coupled at nonlinear level. A combined weakly nonlinear analysis of this case is beyond the scope of this paper. However, there is a particular case where this coupling may be worked out more easily. This is the coupling of the longitudinal modes to the $q=0$ transverse mode. The coupling of a finite $q_o$ mode with a Goldstone mode has been discussed in the literature of liquid crystals electroconvection with homeotropic alignment, \cite{rossberg1996weakly}. This case was shown to be a remakable case of a direct transition to spatio-temporal chaos at onset, due to the nonlinear coupling between the Goldstone mode and the bifurcating mode \cite{rossberg1996weakly}. The scenario was called soft-mode turbulence, and was demonstrated for a stationary periodic instability (i.e. $w_o=0$, $q_o \neq 0$), so it would correspond to our  case with $T_0=0$, for the nonlinear coupling of the soft transverse mode and the longitudinal $q_o$ mode. We are not aware of any study of soft-mode turbulence for a periodic oscillatory mode, but it is again plausible to expect that the dynamics will be no less chaotic. Consequently, we have every indication that the behavior of the system in sufficiently extended 2d regions will generically contain different forms of weak turbulence, possibly in parameter regions where the 1d modes are not yet turbulent.





\section{Application to experiments on epithelial monolayers}\label{sec:discussionandconclusions}

We will now discuss the applicability of the present approach to interpret and even extract quantitative 
information out of existing data, mostly from three series of experiments on MDCK cells under different confinement conditions, but all exhibiting some kind of slow oscillatory dynamics, with characteristic periods of several hours and wavelengths in the range of hundreds of micrometers \cite{Serra-Picamal2012a, Vedula2012, Deforet2014}.

\begin{table*}[t]

\begin{ruledtabular}
\begin{tabular}{cccccccc}
\textrm{$T_{o}$[Pa$\cdot$min]}&  \textrm{$L_{c}$[$\mu$m]} & \textrm{$\eta$[Pa$\cdot$min]}& \textrm{$\xi$[Pa$\cdot$min/$\mu$m$^2$]} & \textrm{$\zeta$[Pa]}& \textrm{$\gamma$[min]}& \textrm{$\rho$[Pa]} &  \textrm{$\nu$[-]} \\
\colrule
$10$ & $10$ & $10^5-10^6$ & $10$ & $10^3$ & $10-100$ & 10 & 10 \\
\end{tabular}
\end{ruledtabular}
\caption{\label{tab:table1}%
Order of magnitud of the model parameters for spreading epithelial monolayers. The parameters $T_{o}$, $L_{c}$, $\eta$ and $\xi$ can be extracted for instance from Ref.~\cite{mercader2017effective}. The coefficient $\zeta$ and $\gamma$ are estimated from Refs. \cite{Roure2005,Weber2012a}, respectively. The coefficient $\rho$ is extracted from Ref.~\cite{Lee2011}. The parameter $\nu$ is estimated to make our model consistent with the rest of parameters and the experimental observations.}
\end{table*}

\subsection{Dynamics of freely spreading epithelia}

The present work was directly motivated by the wound healing in vitro assays described in Ref.~\cite{Serra-Picamal2012a} 
where apparently elastic ultraslow waves were reported. Those experiments study a wound-healing assay, where a wide planar front of the monolayers spreads at an approximately constant speed and with no significant degree of cell proliferation. The data are averaged over the transversal $y$-direction, so the possible structure of the fields along the $y$ coordinate, if present, is averaged out. More specifically, the transverse modes with finite $q_y$ and $q_x=0$ will be present but averaged out in the data, while the transverse modes with finite $q_x$ and $q_y=0$ will generically be present if the instability is sufficiently above threshold to allow them to be unstable. 
This implies that at distances of the order of $L_c$ from the leading edge, the systems is manifestly polarized. Note, however, that since the transversal dimension is much larger than $L_c$, the monolayer could well be polarized in regions further apart from the leading edge, but such polarization be averaged out in the 1d projection. 
The waves reported from those experiments seem perfectly compatible with the waves studied here. The origin of the long period of the waves and their elastic-like phase lag, were indeed puzzling. Several possible explanations have been proposed so far, such as the nonlinear viscoelastic spring model described by \cite{Serra-Picamal2012a} where oscillations reflect sequential fronts of cytoskeletal, or the model by \cite{wavesMarchetti2015}, where no nonlinear elasticity is invoked, but a feedback between local strain, polarization, and contractility is postulated to endow the elastic medium with an effective inertia. Both cases assume that the medium is constitutively elastic, at least partially, and that the origin of the phase-lag between stress and strain-rate is due to the dominance of the elastic relaxation. By contrast, our result shows that the elastic-like phase lag can be entirely associated to the presence of active traction forces, and thus be observed in a medium with a purely viscous constitutive equation. This surprising result is along the lines already suggested in Ref.~\cite{mercader2017effective}, where a purely viscous continuum model for spreading epithelia was shown to explain other apparently elastic behavior such as the emergence of an effective elastic modulus \cite{romaric2015}.
Furthermore, in both models \cite{Serra-Picamal2012a,wavesMarchetti2015} the time scale of the oscillations 
results from internal processes in the medium. 
By contrast, in our model the wave frequency turns out to be extrinsic to the material properties, since it depends essentially on the parameters characterizing the contact forces with the substrate, friction and traction. This is an interesting feature to discriminate the different theories, since these parameters can be changed easily by modifying the properties of the substrate or the molecular complexes that interact with the substrate. 
Finally, in these experiments the waves seem to propagate in the direction of the polarization and not backwards, consistently with our prediction of polar waves.

\subsection{Flow alignment and plithotaxis}

In order to fit the data from Ref. \cite{Serra-Picamal2012a} with our model, we can obtain estimates 
from experimental data of all parameters of our model except the flow alignment coefficient $\nu$ and $\rho$. The values we obtain are listed in Table \ref{tab:table1}. The parameter $\nu$ which couples the flow and the polarity is difficult to measure in living tissues, and is usually not known. We are only aware of values inferred from data for the epithelium of the Drosophila wing, which are negative with $|\nu|$ in the range $3 - 10$ \cite{Aigouy2010}. The use of our model to fit the experimental data gives in our case $\nu \approx 10$ or larger. Values of $|\nu|$ for liquid crystals are only slightly larger than $1$ \cite{de1993physics}. The stronger coupling in the case of tissues may effectively entail an active response of the cells to the environment that can be encoded in this parameter. The value of $\nu$ follows for instance from Eq.~(\ref{eq:vL}) which relates the underlying tissue velocity $V$ with the phase velocity $v_L$ of the waves. Although the presence of transverse modes also travelling along $x$ cannot be ruled out, the presence of the longitudinal mode is clear because of the phase-lag of the stress vs strain rate, which would not be present for purely transverse modes. In any case, for large values of $\nu$ the phase velocity predicted for both types of waves are very similar. 
 
 The fact that $\nu$ is that large has an interesting interpretation. It is known from the hydrodynamics of nematic liquid crystals that this parameter sets the orientation angle $\theta$ of the polarity (director) field with respect to a pure shear flow, such that $\cos{2\theta}=1/\nu$. This simple relation implies that, for 
 $|\nu| \gg 1$ the polarity of cells orients with an angle of $\theta \approx \pi/4$ with the shear. Taking into account that, for a pure viscous shear, the principal stress directions are precisely at this angle, we conclude that the cells in our case tend to reorient along the directions of maximal principal stress, that is, along the axis where the shear vanishes. This tendency has been observed in different situations and has been named plithotaxis \cite{Trepat2009a}. Regardless of whether this response of cells to intercellular stress is an active, regulated process, we find that it is naturally encoded in the parameter $\nu$. 
Consequently, the phenomenon does not need to be seen as an emergent collective property, as it  
can be effectively described as a passive local hydrodynamic coupling between the flow and the polarity.

\subsection{Spreading with lateral confinement}

In Ref.~\cite{Vedula2012} epithelial monolayers migrate along adhesive strips with a controllable width. 
For the most narrow channels, the problem is the closest experimental situation to our specific case of 
$1d$, that is purely longitudinal waves. The modes $q_y$ that are present may be limited to relatively high-$q$, and therefore the transverse instability may be suppressed (not averaged out as in  \cite{Serra-Picamal2012a}). Similarly, the transverse modes with finite $q_x$ and $q_y=0$ will also be suppressed by the boundary conditions on the lateral sidewalls that enforce a fixed orientation of the polarity along them. Consequently, in very narrow channels only longitudinal modes with $q_y=0$ are expected to be relevant. 
For the narrow channels, we associate the contraction-elongation caterpillar-like motion as a signature of our longitudinal waves. In addition, they seem to propagate only in one direction, as predicted by our model.
When the channel width is progressively increased,  
unstable transverse modes are expected to appear and yield the progressively more complex dynamic scenario. The change of behavior for increasing channel width is thus qualitatively and quantitatively consistent with the prediction of our linear analysis. As in all the other cases, the suppressing effect of the treatment blebbistatin is consistent with 
the instability mechanism that we propose for the phenomena. Finally, the complexity of the flow patterns 
observed for wide channels seems to be qualitatively consistent with the scenarios of weak turbulence predicted by our model.

\subsection{Oscillations in totally confined monolayers}

Finally, our model can be used to revisit the experiments of Ref.~\cite{Deforet2014} where the 
monolayers are totally confined in circular islands, but nevertheless exhibit oscillatory collective modes. The fact that the time and length scales are the same as in the other experiments may suggest that the mechanism behind such collective mode could be a similar instability adapted to the confined geometry. Whether our model would yield oscillatory modes in a confined geometry is an open question that we do not address here. However, we can show that the reported linear dependence 
of the oscillation period with the tissue radius $R$ is consistent with our linear dispersion relation analysis. In fact, the oscillation frequency is $\omega_o \sim q T_0 /\xi$, then assuming that the range of wave numbers allowed by the geometry is such that $q > q_o$, then increasing the radius set the most unstable mode available as the minimum $q_{min} \sim R^{-1}$. Consequently, the period will grow linearly with $R$ as reported. 

\subsection{Collective modes and turbulence in epithelia}

In the experiments addressed in the previous section, when the tissue is strongly active, highly disordered flow patterns are observed, often described as noisy \cite{Deforet2014}. Noisy data of local measurements are often reflecting inherent strong fluctuations of the physical variables. In particular in experiments with very large monolayers, simple visual inspection shows an apparently turbulent behavior \cite{PrivateComunication}. Whether this apparent chaos is the manifestation of intrinsic 
noise in the system, or some form of collective modes in a turbulent regime is an interesting open question. 
In this paper we have seen that secondary instabilities after a Hopf bifurcation do generically lead to different
forms of spatio-temporal chaos or weak turbulence, in particular in large systems, so scenario of chaotic collective modes seems plausible. The distinction between the two possibilities is not only of theoretical interest, but may have practical relevance. Indeed, if a turbulent state results from an instability, it can be suppressed or triggered at convenience with just tuning a single parameter across the appropriate boundary in the parameter space. In contrast, if the complex dynamics reflects intrinsic noise, this is virtually impossible to control or regulate. 

\section{Conclusion}
 

In this paper we present a general framework to account for the mechanics of epithelial monolayers. The model is build on the idea that at sufficiently long length and time scales, a continuum hydrodynamic approach can capture a large variety of mechanical aspects of such monolayers, encoding their complex biological regulation in a set of (possibly time-dependent) physical parameters. Our model includes a polarity field
that is not locally aligned with the velocity but coupled to the flow as in nematic hydrodynamics
The contact forces with the substrate contain two contributions, a passive friction force aligned with the velocity, and an active traction force anti-aligned with the polarity. The material exerts also active contractile stresses, and its constitutive equation is taken as that of a viscous fluid for the slow dynamics.
The emergence of an effective elastic modulus, for instance, has been shown to be possible in an active viscous fluid  \cite{mercader2017effective}. Here we show that the presence of waves with an elastic-like phase lag between stress and strain rate is not necessarily a signature of elasticity, but can occur in viscous fluids with active tractions. It is worth remarking that both the direct observation of the relative cell movements, and the arguments based on the turnover time scales of processes that control the short-time elasticity of the medium suggests that the rheology of a spreading monolayer should be expected to be that of a viscous fluid. Indeed, here we show that all long-time observations in spreading epithelia fit well with the description in terms of an active viscous polar fluid.

Our framework provides insights into the physics of collective cell migration. The idea 
that interacting cells collectively set the local stress environment and the motion of individual cells, is incorporated in our physical picture through the nonlocal character of the interaction, which establishes that the flow velocity at a given point is determined by an integration over a region of 
the size of the friction length. Similarly, the tendency of cells to align along the direction of principal stresses (plithotaxis), appears in our framework as a consequence of the large values of the flow alignment coefficient $\nu$, which are obtained independently to fit observations on the propagating waves. This parameter turns to be crucial to explain the instability leading to longitudinal waves, which are distinguished in the experiments by the elastic-like phase lag. 

The test of the quantitative predictions of this type of continuum model is not simple because of the difficulty to determine the model parameters and also because these may be changing with time due to the ongoing biological regulation, that may change for instance the properties that are encoded in those parameters. Nevertheless, we remark that  
our model is predictive also in qualitative aspects.
For instance, we predict that the stress waves must be polar, in the sense that they should only propagate along the polarization of the medium and not backwards, as opposed to elastic waves. We also predict that the wave frequency, as well as phase lag are essentially determined by contact forces (friction and traction), which are in principle easier to control in experiments, rather than the material parameters. 

In addition, we have pursued our study of waves into the nonlinear regime, and have shown that different forms of weak turbulence are generically present in the nonlinear waves that emerge in our model. We speculate that this chaotic dynamics of the waves may be at the root of the noisy dynamics of tissues. In particular, experiments in very large spreading monolayers exhibit what could be loosely described as turbulence \cite{PrivateComunication}. Within this picture, the dynamic disorder could well be an expression of highly unstable collective modes, and not a signature of intrinsic noise. This idea could be tested also qualitatively by observing sudden changes between regular and irregular collective behavior by changing a single parameter, which could be interpreted as the transition from a chaotic to a regular regions in the parameter space. 

The model is not expected to apply to epithelial tissues that are not moving on a substrate \cite{harris2012}.
It is unclear to what extent the model can be adapted to situations where there is no global flow, such as in the fully confined experiments of Ref.~\cite{Deforet2014}. However, for the slow flow dynamics of epithelia on substrates, the physical scenario here unveiled is expected to be generic. Indeed, 
 even though the explicit exact calculation here presented for both linear and nonlinear dynamics refer to a specific choice of the terms included in the model, which has been kept as simple as possible, our central results are robust and do not depend on the model details. The model could be enriched with more parameters, and new physical ingredients, such as effects of cell division or short time elasticity. However, from the generic nature of the linear and nonlinear analysis here discussed, which relies to a large extent on symmetries, it is expected that two basic results are robust to changes in model details, namely, the mechanism that controls the traction-driven instability of an active viscous polar fluid, leading to polar nonlinear travelling waves, and the nonlinear instabilities that lead generically to weak turbulence scenarios. The need of at least two sources of activity, and a coupling between polarity and flow seem also well established. We expect that further experimental inquiry will eventually test the ideas here developed and clarify the appropriate mechanical framework for a continuum description for collective cell migration in tissues.


We thank R. Alert, F. Graner, V. Hakim, J.-F. Joanny, P. Marcq, J. Prost, X. Trepat, R. Vincent and S. Yabunaka for useful discussions. 
We acknowledge financial support from MINECO under projects FIS2013-41144-P and FIS2016-78507-C2-2-P
from Generalitat de Catalunya under project 2014-SGR-878. CBM also acknowledges a FPU fellowship from the Spanish Government.

\section{Appendix A}

Here we provide details of the weakly nonlinear analysis leading to the CGLE in our physical model for the
case of $1d$ longitudinal modes. We use the corresponding physical model given by
Eqs.~(\ref{eqn::forcebalance1}-\ref{eqn::molecularfield1}) and assume a system of units such that $\gamma=\eta=\xi=1$.

%

Close to threshold and expanding the linear growth rate around its maximum at $q_o$, we get 
\begin{eqnarray}
\text{Re}[\omega(q)]&=&\frac{2\nu \left(\zeta_{o}^{L}-\zeta\right)}{\left(4+L_{c}\sqrt{8+2\rho\nu^2}\right)}\nonumber
\\&-&\frac{4\sqrt{2}L_{c}^2(4+\nu^2\rho)}{2\sqrt{2}+L_{c}\sqrt{4+\nu^2\rho}}(q-q_{o})^2+\dots
\label{append:realgrowthrate}
\\ \frac{\text{Im}[\omega(q)]}{T_{0} q}&=&1+\frac{\nu L_{c}\sqrt{4+\nu^2\rho} }{2L_{c}\sqrt{4+\nu^2\rho}+4\sqrt{2}}+ \dots
\label{append:imaggrowthrate}
\end{eqnarray}
with
\begin{equation}
\zeta_o^{L} = - \frac{2}{\nu} \left( L_c + \sqrt{2+\frac{\rho\nu^2}{2}} \right)^2
\end{equation}
The weakly nonlinear analysis is a formal expansion on the small parameter 
$\epsilon \equiv (\zeta-\zeta_{o}^L)/\zeta_{o}^L$ that measures the normalized distance to threshold. We will refer to $\epsilon > 0$ as the system being (slighly) above threshold. Then, a narrow band of size $|q-q_o| \sim \sqrt{\epsilon}$  are unstable but nearly marginal, since for them $Re[\omega] \sim \epsilon$. On the contrary, modes with 
$|q-q_o| \sim {\epsilon}^0$ will relax much faster, with $Re[\omega] \sim \epsilon^0$. 
 Accordingly, long wavelength spatial modulations of order $\epsilon^{-1/2}$ with slow relaxation times of order $\epsilon^{-1}$ are expected to dominate the dynamics, and slave the rest of (fast) modes. This separation of time and length scales is at the root of the universal description in terms of an amplitude equation for the envelope of the bifurcating mode. 


Consequently, in general the perturbations about the reference ordered state can be expressed as a superposition of plane waves with wavenumber multiples of $q_{o}$ and phase velocity $v_{o}=-\text{Im}[\omega(q_{o})]/q_{o}$ plus an envelope wave with slow spatio-temporal dynamics. In our particular case the most general solution reads     
\begin{eqnarray}
p&=&1+\sum_{n=1}^{\infty}\sum_{m=-n}^{n}\epsilon^{\frac{n}{2}}p_{nm}(X,T)e^{i m q_{c}(x-v_{c}t)}\label{eq:CGLEansatz1}
\\ v&=&T_{0}+\sum_{n=1}^{\infty}\sum_{m=-n}^{n}\epsilon^{\frac{n}{2}}v_{nm}(X,T)e^{i m q_{c}(x-v_{c}t)}
\\ \sigma&=&-\frac{\zeta}{2}+\sum_{n=1}^{\infty}\sum_{m=-n}^{n}\epsilon^{\frac{n}{2}}\sigma_{nm}(X,T)e^{i m q_{c}(x-v_{c}t)}
\\ V_g&=&\sum_{n=0}^{\infty}\epsilon^{\frac{n}{2}}V_{n}\label{eq:CGLEansatz4}
\end{eqnarray}
where the envelope waves of the corresponding physical fields $p_{nm}$, $v_{nm}$ and $\sigma_{nm}$ are functions of the long spatial variable $X\equiv\epsilon^{1/2}(x+ V_g t)$ and the slow temporal variable $T\equiv\epsilon t$.  $V_g$ is the travelling speed of the wave envelope (the group velocity) and in general is a power series in $\epsilon^{1/2}$, whose coefficients are treated as unknowns. 

The physical solution valid near the vicinity of the transition (\ref{eq:CGLEansatz1}-\ref{eq:CGLEansatz4}) is inserted into the PDE's (\ref{eqn::forcebalance1}-\ref{eqn::molecularfield1}) and the different terms are sorted in powers of $\epsilon^{1/2}$. Note that the zeroth order leads to the ordered uniform solution. The first order turns into an undetermined set of linear equations for $\sigma^{m}_{1}$, $v^{m}_{1}$ and $p^{m}_{1}$. As an example these coefficients solved as a function of the amplitude $p_{1}^{1}$ read
\begin{eqnarray}
v_{11}&=&\frac{2 q_{o} \left(i L_{c}^2 q_{o}+v_{o}-T_{0}\right)+4 i}{\nu q_{o}}p_{11}
\end{eqnarray}
Similarly the second order can also be arrange into a linear set of equations for the coefficients $\sigma_{2m}$, $v_{2m}$, $p_{2m}$ and $V_{1}$. Through a solvability condition these second order coefficients are connected to $p_{1m}$. These conditions are often used to construct self-consistent solutions through perturbative analysis. Lastly, the third order solvability condition is analogous to the Complex Ginzburg-Landau equation. Thus after rescaling the variables conveniently, it can be expressed in the form
\begin{equation}
\partial_T p_{11}=p_{11}-s(1+i c)|p_{11}|^2p_{11}+(1+i b) \partial^2_{X} p_{11},\label{eq:amplitudeequation}
\end{equation}

being $s$, $c$ and $b$ parameters that in general depend on the details of the mechanical properties of the system. The coefficient $s$ is either $\pm 1$ and it controls whether the transition is continuous or discontinuous. With respect to Eq.~(\ref{CGLE}), we have replaced the variable $p_{11}$ by $A$. 

The longitudinal mechanical transition in our system is oscillatory and periodic, meaning that the critical wavenumber $q_{o}$ is finite and also the travelling speed of the perturbations $v_{o}$. As a result, the coefficients of our amplitude equation (\ref{eq:amplitudeequation}) are complex. Only in the particular case of null active traction forces (i.e. $T_{0}=0$) both coefficients $b$ and $c$ vanishes, reducing Eq.~(\ref{eq:amplitudeequation}) to the Real Ginzburg-Landau equation for $s=1$. This equation is purely relaxational, that is, it exist a Lyapunov function that is maximised over time. Except for some particular cases no such Lyapunov functional can be constructed for the CGLE, giving rise to a richer phenomenology of dynamical states: from travelling coherent states to different forms of spatiotemporal chaotic states.



For completeness, we present the analytical form of the coefficients $s$, $c$ and $b$ as
\begin{eqnarray}
b&=&\frac{F}{2}\frac{(4-3G\sqrt{8+2H})}{\sqrt{G}\sqrt[4]{8+2H} (8+G\sqrt{8+2H})^2},
\\ c&=&-\frac{f_{2}f_{4}-f_{1}f_{3}}{|f_{1}f_{2}+f_{3}f_{4}|},
\\ s&=&-{\rm Sign}\left(\frac{f_{1}f_{2}+f_{3}f_{4}}{f_{2}^2+f_{3}^2}\right),
\end{eqnarray}
where the function Sign returns the sign of its argument. The functions $f_{1}$, $f_{2}$, $f_{3}$ and $f_{4}$ read
\begin{widetext}
\begin{eqnarray}
f_{1}&=& 16 \Big\{2 G^5 (H+4)^2 \Big[27 \nu^2-42 \nu-8\Big]+16 \sqrt{2} (H+4)^{3/2} \Big[H^2 \nu^2+H \nu-20\Big]\nonumber
\\&&-2 \sqrt{2} G^4 (H+4)^{3/2} \Big[3 (5 H-36) \nu^2+(13 H+228) \nu+2(H+52)\Big]\nonumber
   \\&&-G^3 (H+4) \Big[H^2 (65 \nu^2-104 \nu-4)+4 H (135 \nu^2-52 \nu+52)-16 (27 \nu^2-90 \nu-128)\Big]\nonumber
   \\&&+2 \sqrt{2} G^2 \sqrt{H+4} \Big[2 H^3 \nu (\nu+11)-3 H^2 (35 \nu^2-96 \nu+4)-12 H (51 \nu^2-62 \nu+44)-32 (3 \nu+76)\Big]\nonumber
   \\&&+8 G (H+4) \Big[H (H^2+6 H-48) \nu^2+(23 H^2+100 H+48)\nu-4 (9 H+88)\Big]\Big\}\nonumber
   \\ && -F^2 (H+4) \nu^2 \Big\{\sqrt{2} G^2 \sqrt{H+4} (\nu+2) \Big[ H (\nu+2)+52 \nu-88\Big]\nonumber
   \\&&-8 G \Big[4 (H+4) \nu^2+(13 H+4) \nu+2 (5H+68)\Big]-96 \sqrt{2} \sqrt{H+4} (\nu+4)\Big\}
  \\f_{2}&=&36 G (H+4)^{2} \nu^2 \left(G \sqrt{H+4}+2 \sqrt{2}\right)^2
\\ f_{3}&=&-24 \sqrt[4]{2} F \sqrt{G} (H+4)^{7/4} \nu^3 \left(G \sqrt{H+4}+2 \sqrt{2}\right)
\\ f_{4}&=&3 \sqrt[4]{2} F \sqrt{G} \sqrt[4]{H+4} \nu  \Big\{-4 G^3 (H+4) \Big[(9 H+68) \nu^2+4 (H-12) \nu-28 (H+4)\Big]\nonumber
\\&&+\sqrt{2} G^2 \sqrt{H+4} \Big[5 H^2 \left(3 \nu^2+10 \nu+8\right)+32 H \left(\nu^2+7 \nu+26\right)-368 \nu^2+608 \nu+2688\Big]\nonumber
\\&&+4 G (H+4)\Big[\left(2 H^2+63 H+44\right) \nu^2+4 \left(H^2+8 H+28\right) \nu+96 (H+7)\Big]\nonumber
\\&&+16 \sqrt{2} (H+4)^{3/2} \left(3 H \nu^2+(7-4 H) \nu+28\right)\Big\}
\end{eqnarray}
\end{widetext}
and only depends on the dimensionless physical parameters $F=T_{0}\gamma/\xi L_{\eta}$, $H=\rho\gamma \nu^2/\eta$ and $G=L_{c}/L_{\eta}$.

\bibliography{bib-tesis-complet}

\end{document}